\documentclass{article}
\usepackage{spconf,amsmath,epsfig}
\usepackage{soul,color}
\usepackage{adjustbox}
\usepackage{amsmath}
\usepackage{amssymb}
\usepackage{amsthm}
\usepackage{algorithm}
\usepackage{algorithmic}
\usepackage{float}
\usepackage{graphicx}

\usepackage{hyperref}
\hypersetup{
    unicode=false,          % non-Latin characters in Acrobat’s bookmarks
    pdftoolbar=true,        % show Acrobat’s toolbar?
    pdfmenubar=true,        % show Acrobat’s menu?
    pdffitwindow=false,     % window fit to page when opened
    pdfstartview={FitH},    % fits the width of the page to the window
    pdftitle={},    % title
    pdfauthor={},     % author
    pdfsubject={},   % subject of the document
    pdfcreator={},   % creator of the document
    pdfproducer={}, % producer of the document
    pdfkeywords={Forward}{inverse}{modelling}{Gaussian process}{kernels}, % list of keywords
    pdfnewwindow=true,      % links in new window
    colorlinks=true,       % false: boxed links; true: colored links
    linkcolor={blue},          % color of internal links (change box color with link)
    citecolor=blue,        % color of links to bibliography
    filecolor=blue,      % color of file links
    urlcolor=blue           % color of external links
}
\usepackage{color}

\title{Retrieval of Case 2 Water Quality Parameters with Machine Learning } % shorter is better (more citable and more memorable)
\name{Ana B. Ruescas$^1$, Gonzalo Mateo-Garc{\'i}a$^1$, Gustau Camps-Valls$^1$ and Martin Hieronymi$^2$\thanks{The research was funded by the European Research Council (ERC) under the ERC-CoG-2014 SEDAL project (grant agreement 647423), and the Spanish Ministry of Economy and Competitiveness (MINECO) through the project TIN2015-64210-R. Especial thanks to Carsten Brockmann and the Case2eXtreme project team (funded by ESA).}}{}
\address{$^1$Image Processing Laboratory (IPL), Universitat de Val\`encia, Spain\\
$^2$ Institute of Coastal Research, Helmholtz-Zentrum Geesthacht, Germany}

\begin{document}
\onecolumn
%\ninept
%

\maketitle
\vspace*{-8mm}
\begin{abstract}
Water quality parameters are derived applying several machine learning regression methods on the Case2eXtreme dataset (C2X). The used data are based on Hydrolight in-water radiative transfer simulations at Sentinel-3 OLCI wavebands, and the application is done exclusively for absorbing waters with high concentrations of coloured dissolved organic matter (CDOM). The regression approaches are: regularized linear, random forest, Kernel ridge, Gaussian process and support vector regressors. The validation is made with and an independent simulation dataset. A comparison with the OLCI Neural Network Swarm (ONSS) is made as well. The best approached is applied to a sample scene and compared with the standard OLCI product delivered by EUMETSAT/ESA. 
\footnote{Paper published in IGARSS 2018 IEEE, DOI 10.1109/IGARSS.2018.8518810}
\end{abstract}
\section{Introduction}
\label{sec:intro}
\vspace*{-2mm}
Case 2 absorbing waters, mainly located in the temperate and cold regions of the boreal zone -in Europe waters in Finland, Sweden and Estonia-, are challenge areas for Earth Observation water quality (WQ) estimation. Even if a widely studied problem, current methods often yield to large errors and uncertainties. These waters typically have fairly low total suspended matter (TSM) and chlorophyll\_a (Chl-a) contents, even though some cases of ``black lakes'' with high Chl-a and TSM values have been reported. The high dissolved organic matter concentration in these humic waters increases attenuation of light in the blue and green regions of the spectrum, and consequently decreases reflectance in the short wavelengths ~\cite{Kallio183}. Remote sensing techniques are used to retrieve the colored dissolved organic matter (CDOM) and derive the dissolve organic carbon (DOC), which plays a significant role in the carbon and energy cycle of lakes, and affect the quality of drinking water too. The photosynthetic pigment Chl-a is a key indicator of phytoplankton biomass, and knowing its concentration in water is essential for monitoring its quality. Chl-a is usually consider as the main parameter, yet CDOM is the more uncertain ocean color (OC) product. Information on water quality is then used in official directives like the Water Framework Directive (WFD) of the European Commission.
Many of the algorithms developed to extract the three mentioned WQ parameters (CDOM, Chl-a, TSM) from remote sensing observations are based on empirical or semi-empirical methods, which are very accurate locally, but fail when the spatio-temporal range is surpassed~\cite{ATTILA2013138}. The objective of this study is to test the applicability for WQ retrieval of methods coming from the machine learning field, based both on linear and non-linear regression algorithms. The inputs used for training the models are the data from the C2X project database, which is based on simulated remote sensing reflectance ($R_{rs}$). We will explore the effectiveness of standard machine learning regression algorithms to map reflectances into WQ parameters. Namely, we will resort to neural networks, random forests, and kernel regression algorithms. 
In the present study, we use Hydrolight simulation datasets developed in the framework of the \textit{Case2eXtreme} (C2X) project\footnote{\url{http://seom.esa.int/page_project014.php}}~\cite{Hieronymi2016}. The simulations selected here represent different conditions of the medium to high absorbing waters, with cases similar to the northern coastal Baltic Sea waters and many inland waters. These are Case 2 waters with high amounts of CDOM, and with great seasonal variation of Chl-a concentrations from the spring blooms ($10-120 mg m^{-3}$) to the summer minimum ($1-3 mg m^{-3}$) ~\cite{ATTILA2013138}. During the central summer months, surface cyanobacteria blooms appear regularly (mean concentration of $5-30 mg m^{-3}$). High absorption by CDOM distorts the Chl-a estimations at wavelengths $<500 nm$, which are the ones typically used in global Chl-a algorithms. The estimation of suspended solids is also important since it is related to water turbidity, and used as a parameter for the estimation of the water quality too. 
\vspace*{-2mm}

% \textcolor{red}{The remainder of the paper is organized as follows.
% \S2 describes the data collection and characteristics, 
% while \S3 briefly reviews the methods used in this work.
% \S4 gives an empirical evidence of performance of the proposed methods in comparison to standard bioptical models for the particular dataset used.
% We conclude in \S5 with some remarks and an outline future work.}

\section{Methods and application}
\label{sec:ML}
\vspace*{-2mm}
\subsection{Data collection}
\label{sec:DATA}
\vspace*{-2mm}
Highly absorbing or black waters are characterized by very low water-leaving radiance over the visible range. The maximum of $R_{rs}$ is typically $< 0.005 sr^{-1}$ and located between 550 and 605 nm for Case-2 absorbing (C2A) cases. For extremely Case-2 absorbing waters (C2AX), the maximum shifts towards the red spectral range $> 600 nm$. 
We have selected all data classified as C2A and C2AX from the original \textit{Case2eXtreme} dataset and we filtered it to cases with most widespread Chl-a-specific spectral absorption, i.e. not related to cyanobacteria. In total we have obtained 5570 data that we use to train the models, plus independent $\approx 1800$ spectra for validation purposes. The spectral range is the one corresponding to the Sentinel-3 Ocean and Land Color Instrument (OLCI) configuration \footnote{\url{https://sentinel.esa.int/web/sentinel/technical-guides/sentinel-3-olci}}, with a total of 15 bands: 400, 412.5, 442.5, 490, 510, 560, 620, 665, 673.75, 681.25, 708.75, 753.75, 778.75, 865,885 nm. The CDOM (440 nm) absorption coefficient has a range between $0.098$ and $20 m^{-1}$; while the Chl-a content range rises until $200 mg m^{-3}$.The TSM ranges between $0.02-10 g m^{-3}$. The broad water conditions of this dataset will serve as test for the global applicability of the proposed approaches.
\vspace*{-2mm}
\subsection{Traditional approaches and ML methods}
\label{sec:ratios}
\vspace*{-2mm}
Empirical or semi-empirical algorithms used for the estimation of WQ parameters in C2A(X) waters are usually based on band-ratio algorithms. Those algorithms are mostly based on airborne and field measurements with negligible or only small atmospheric influence. Important wavelength regions for the band ratio algorithms are: for CDOM retrieval between 400-600 nm taking 660-720 nm as reference. %Following~\cite{Kallio183, Brezonik2015199}, CDOM can be estimated by a ratio of reflectance at wavelength $> 600$ nm to reflectance in the 400-550 nm range. This ratio is valid in a wide range of water constituent combinations.~\cite{Alikas14,Kallio183} used {\em in situ} measured data in Finland, and derived model coefficients that work well for CDOM and TSM. 
Chl-a algorithms based on band ratios or semi-analytical approaches are also centered in a few bands of the spectrum on the blue and the green spectral regions ~\cite{LIGI201757}. The TSM is usually estimated using a single band at 709 nm and not a ratio. Physically-based analytical approaches that use the full spectra are becoming more popular, but they are still computationally expensive. Neural network (NN) methods like the ones developed for MERIS \cite{Doerffer07} and OLCI \cite{Hiero17} require of high quality input data and very good inversion models. The advantage of these more sophisticated methods is that they produce reasonable results and are not limited spatio-temporally.
%\subsection{Machine learning approaches}\label{sec:MLA}
Machine learning approaches constitute an alternative to state-of-the-art models such as NN, or other well-established models, such as the NASA OC4. With the improvement of computer resources in computational times and robustness, more theoretical and experimental studies are developed to assess performance of these methods in a wide range of situations. In this exercise we use five machine learning (ML) algorithms for linear and non-linear regression: the (multivariate) linear regression (RLR), random forest regression (RFR)~\cite{Breiman85}, the kernel ridge regression (KRR)~\cite{Shawetaylor04}, the Gaussian process regression (GPR)~\cite{Rasmussen06} and the support vector machine regression (SVR). The code reproducing the results of this paper can be found in \url{https://github.com/IPL-UV/mlregocean}. In addition, an operational Matlab toolbox implementing all machine learning methods is available at \url{https://github.com/IPL-UV/simpleR}. % ;)
\vspace*{-2mm}

\subsection{An application example}
\vspace*{-2mm}
The experiment consists of retrieval of the three basic WQ parameters (Chl-a, TSM and CDOM) using the subset of the C2X dataset (see Section \ref{sec:DATA}). Here, the $R_{rs}$ is the ratio of water-leaving radiance to downwelling irradiance above the sea surface. It refers to clear atmosphere with Sun at zenith and viewing angle exactly perpendicular. The OLCI spectral bands used are the common set found in many of the ocean color sensors used for WQ retrievals, and the one used by ESA in the official ocean product processing of OLCI data. Statistics used to check the validity of the methods are: the coefficient of determination ($R^2$), the bias, the mean absolute error (MAE), the Pearson's coefficient of correlation (R), and the root mean squared error (RMSE) in absolute (abs) quantities.
%%%%%%%%%%%%%%%%%%%%%%%%%%%%%%%%%%%%%%%%%%%%%%%%%%%%%%%
\begin{figure}[!ht]
\centering\includegraphics[width=0.7\linewidth]{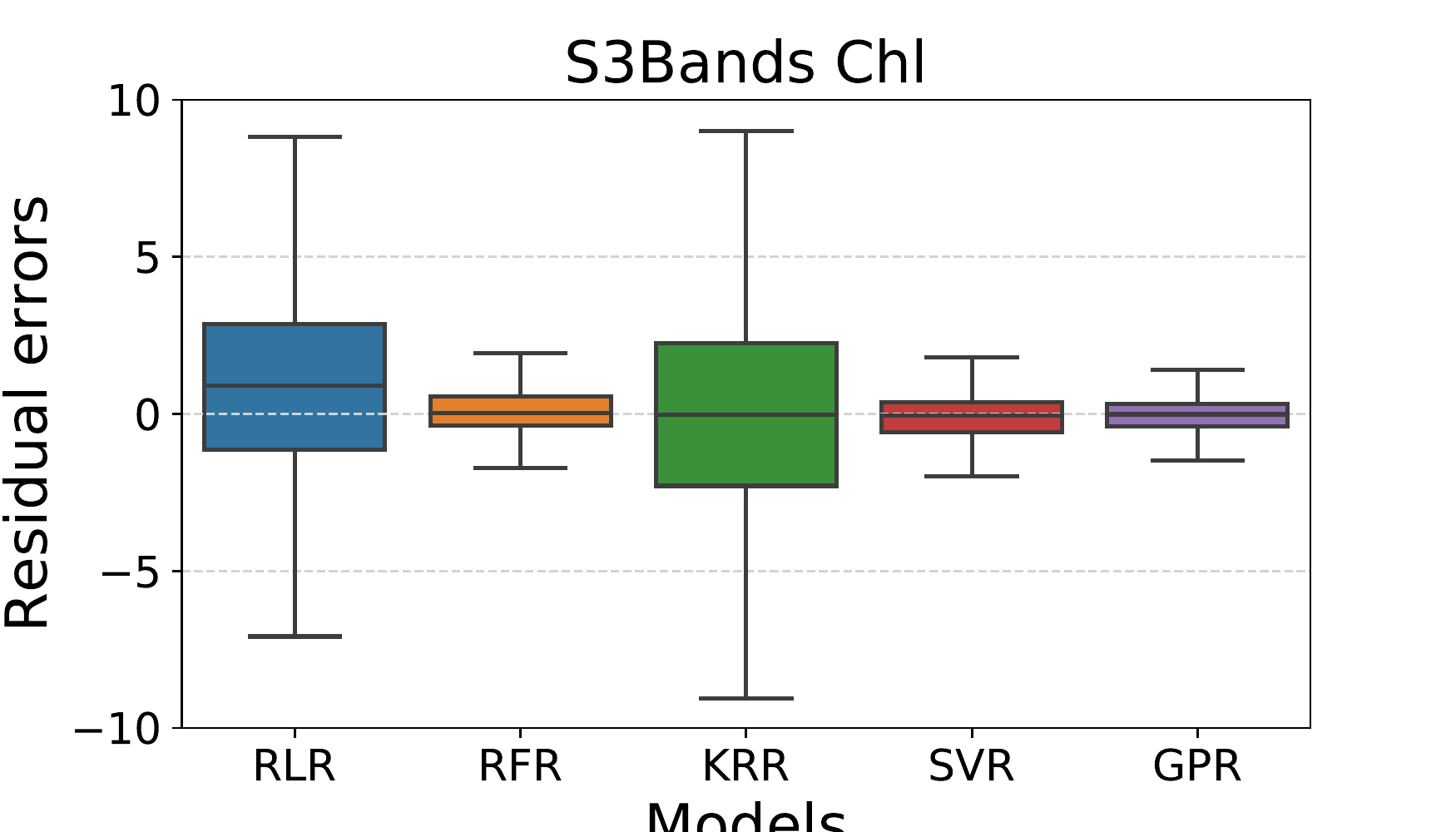}\\
\centering\includegraphics[width=0.7\linewidth]{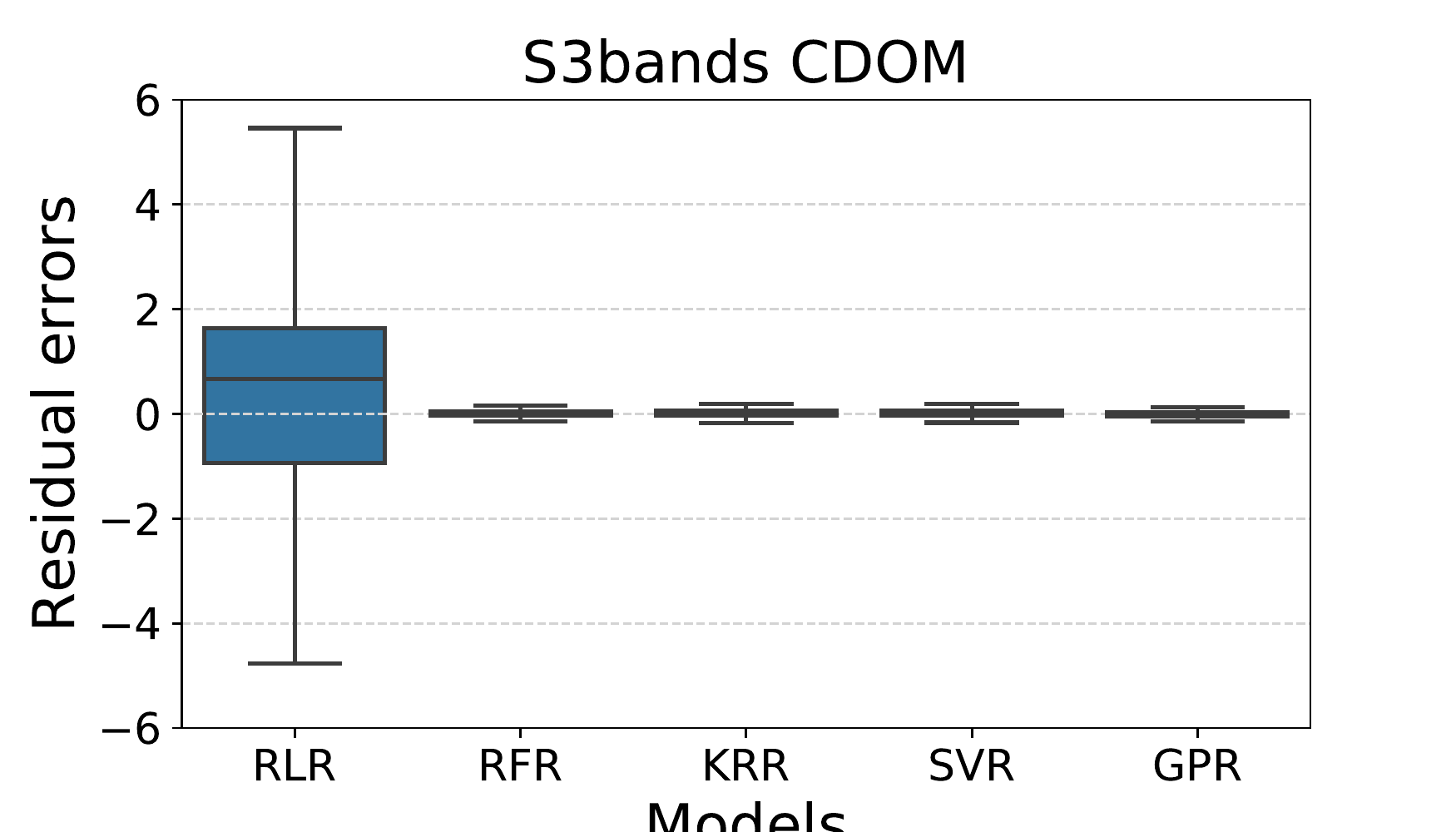}\\
\centering\includegraphics[width=0.7\linewidth]{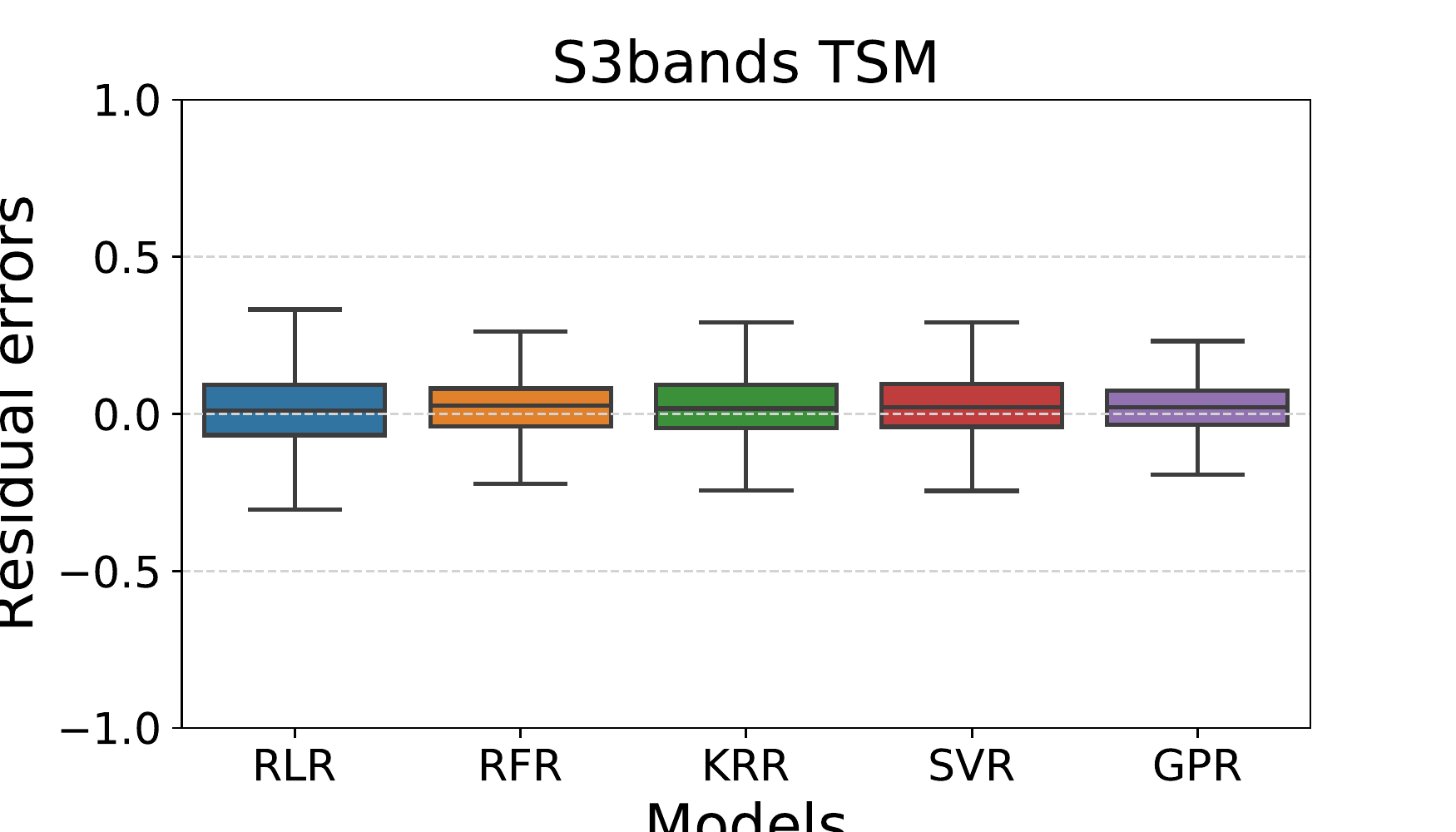}\\
\vspace*{-2mm}
\caption{Application of model on test data: boxplots of the residuals per model and parameter}\label{Fig1}
\end{figure}
\vspace*{-5mm}

\section{Experimental Results}
\label{sec:simpleR}
\vspace*{-2mm}
\subsection{Numerical comparison} 
\vspace*{-2mm}
Fifteen OLCI bands are used as input for the five ML models. Other type of inputs, like band ratios \cite{Ruescas18}, are not included to test the global applicability of the methods, since the use of band ratios have usually a local component difficult to extrapolate. The Table \ref{Tab:stats} offers an overview of the metrics. The analysis by method and by parameter shows that RLR only works well with TSM. The RFR seems to work quite well for CDOM and TSM, with high correlation coefficients and low errors, but not so ideal for Chl-a. KRR and SVR have a good performance for Chl-a estimation, while the performance of the GPR approach is not as good as expected, in fact is surprisingly low. The GPR gives the best results when retrieving CDOM. The TSM variable seems to be less complex to retrieve, and all five methods give very good results, being KRR and the SVR the best ones. Figure \ref{Fig1} shows the distribution of the residual errors comparing the five models side by side for the three WQ parameter. The Chl-a estimation has larger errors than the other two (see change of the scale at the y axis). This is especially noticeable for RLR and KRR, being these approaches less robust to outliers. Both methods improve for CDOM and TSM, with in general better results except for the RLR in CDOM.
\vspace*{-2.5mm}

\begin{table}[!t]
\caption{Numerical results obtained obtained with the machine learning methods. Several scores are shown: coefficient of determination (R$^2$), bias, mean absolute error (MAE), Pearson's correlation coefficient (R) and absolute root-mean-square error (RMSE).}
\centering
\small
\setlength{\tabcolsep}{3pt}
\begin{tabular}{|l|c|c|c|c|c|c|}
\hline
\textbf{Models} & \textbf{R$^2$} & \textbf{bias} & \textbf{MAE} & \textbf{R} & \textbf{RMSE} \\
\hline
{\bf S3 bands, Chl} & & & & & \\
\hline
RLR &0.553&0.021 &5.547 &0.746 & 16.955\\
RFR &0.750 & -0.182& 3.199& 0.866& 12.656\\
KRR  &0.730&-0.405&4.971& 0.855&13.167 \\
GPR &0.438&-0.894&4.356 &0.711 &19.001 \\
SVR &0.698&-1.495&3.098 &0.842 &13.922 \\
\hline
\textbf{S3 bands, CDOM} & & & & & \\
\hline
RLR &0.605 & 0.001& 1.781& 0.781& 2.567\\
RFR &0.991 & 0.005& 0.144& 0.995& 0.372\\
KRR  &0.983 & 0.019&0.224&0.991& 0.537\\
GPR &0.996& 0.008&0.113& 0.998& 0.267\\
SVR &0.892&-0.245&0.683 &0.952 &1.345 \\
\hline
\textbf{S3 bands, TSM} & & & & & \\
\hline
RLR &0.962& 0.020& 0.104& 0.981& 0.149\\
RFR &0.978 & 0.022& 0.083& 0.989& 0.114\\
KRR  &0.971&0.025 & 0.088& 0.986& 0.130\\
GPR &0.891& 0.012& 0.085& 0.945 & 0.252\\
SVR &0.971&0.029&0.088 &0.986 &0.130 \\
\hline
\end{tabular}
\label{Tab:stats}
\vspace*{-7mm}
\end{table}
%\vspace*{-5mm}

\subsection{Analysis of the models}
\begin{figure}[!ht]
\centering\includegraphics[width=0.9\linewidth]{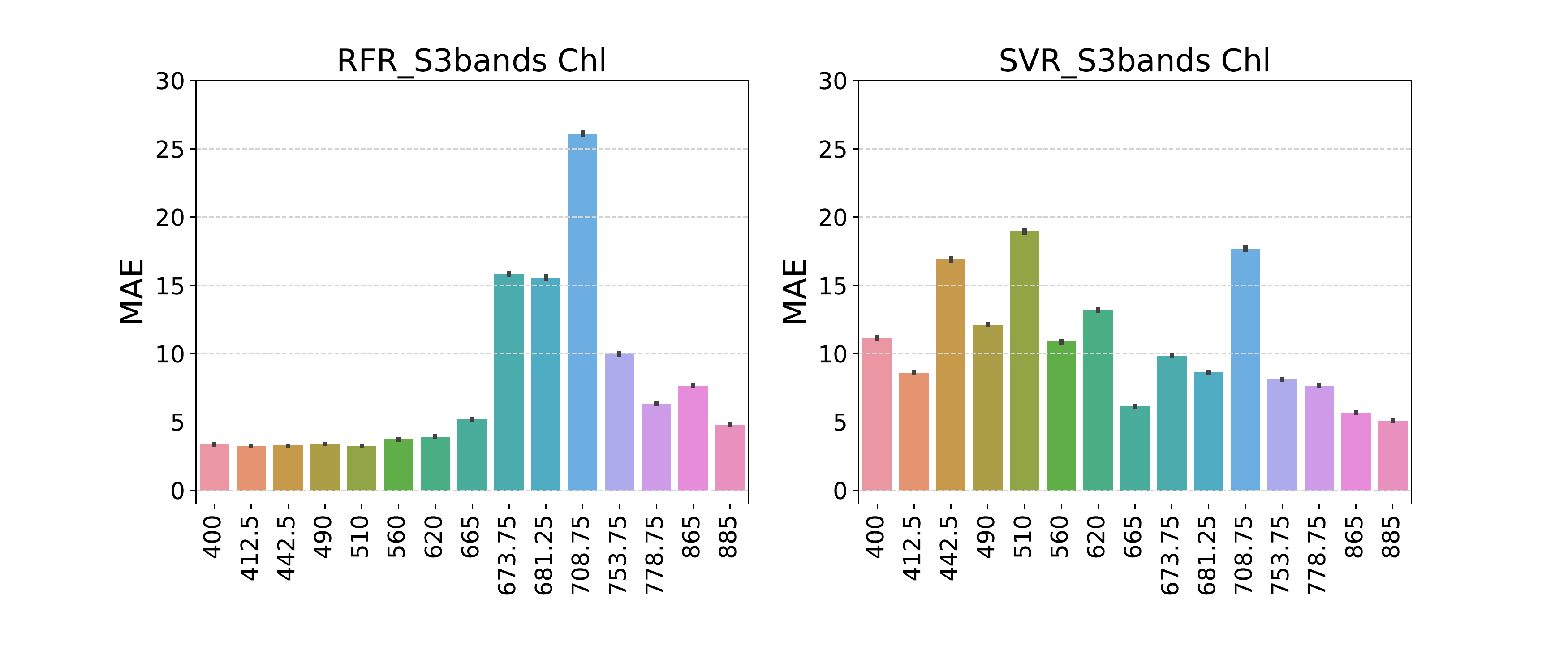}\\
\centering\includegraphics[width=0.9\linewidth]{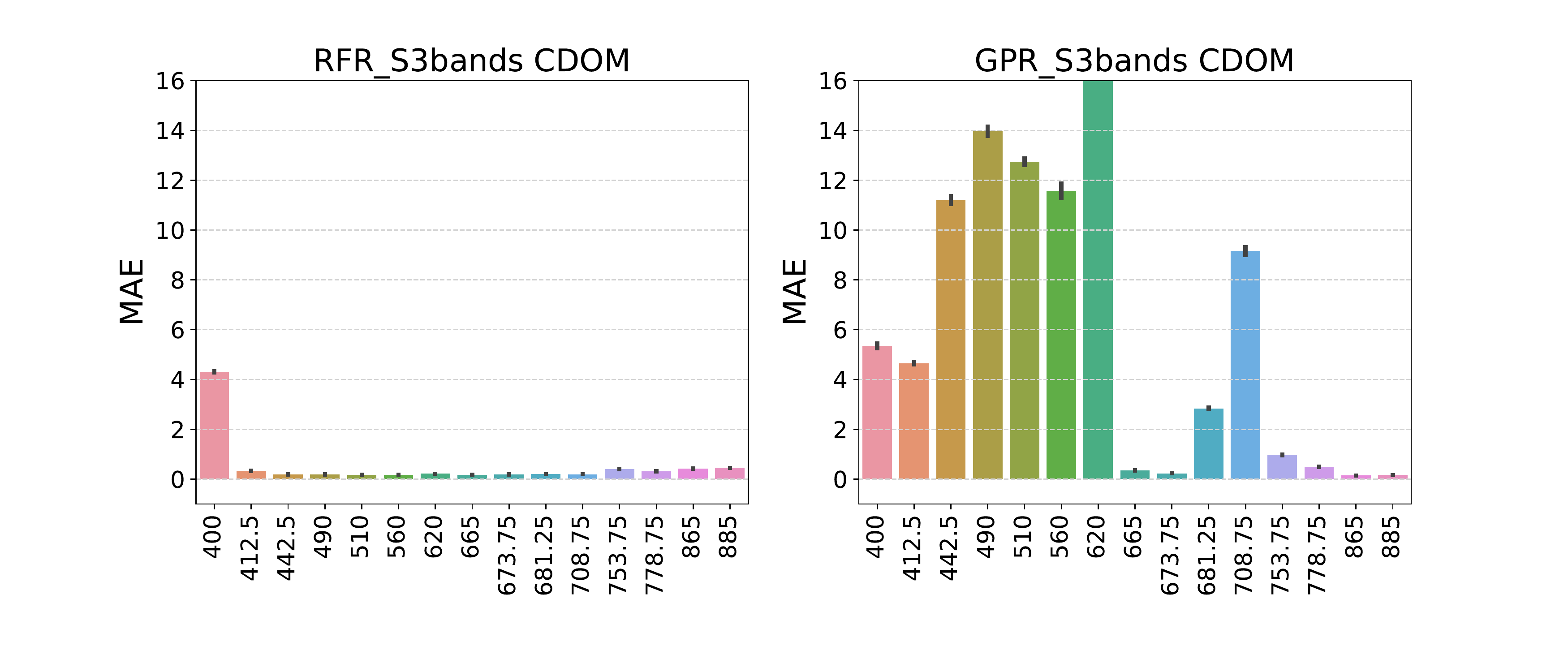}\\
\centering\includegraphics[width=0.9\linewidth]{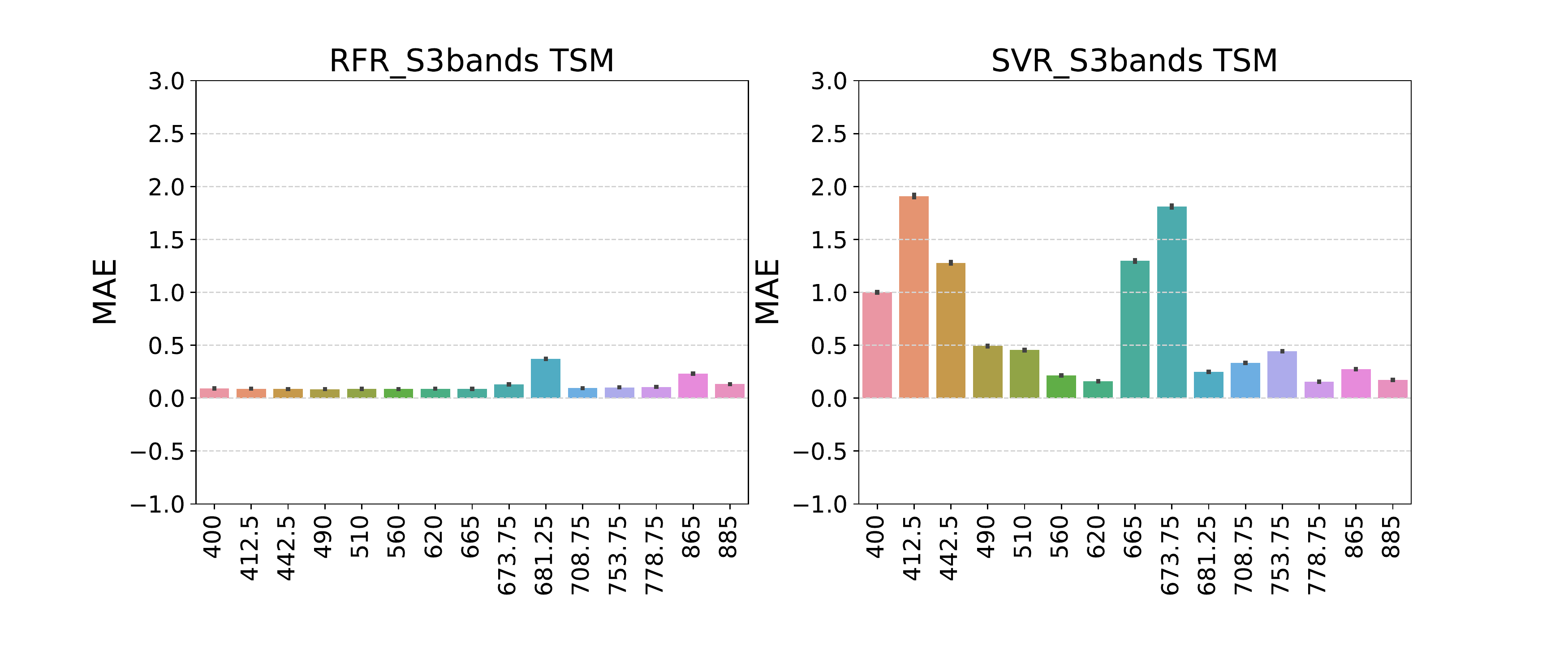}
\vspace*{-5mm}
\caption{Application of model on test data: permutation plots}\label{Fig2}
\end{figure}

%\vspace*{-5mm}
The analysis of the models is made using the permutation plots shown in Figure~\ref{Fig2}. The permutation test measures the relevance of each input, in this case the bands -they are shown as bars of different colors- in the models by accounting the error produced when that input is removed from the set. Only the permutation plots of the best models are shown in Figure~\ref{Fig2} (Chl-a, CDOM and TSM from top to bottom). The RFR model is one of the simplest approaches, and still gives fair results for all three cases. For Chl-a the bands with more weight are the ones in the red part of the spectrum (673 to 709 nm), which are effectively the ones usually taken for the empirical band ratio algorithms in Case 2 waters. The SVR permutation plot for Chl-a tells a slightly different story, and the bands in the blue and green have more relevance here, giving to the red part of the spectrum still certain role. The CDOM RFR plot shows the maximum weight of the 400 nm blue band, which is also expected because the absorption maximum of CDOM takes place in these spectral ranges (400-440 nm). The contrast of this band with the spectrum in the red (620-709 nm) are the inputs used in CDOM band-ratio algorithms in many studies. The TSM permutation plot for RFR shows the importance of the 680 nm and $>800$ nm bands for the determination of suspended solids. The most relevant bands for SVR method for TSM are again the red range and its contrast with the blue bands, combination used also in many band-ratio empirical algorithms \cite{LIGI201757}.
\vspace*{-2mm}

\section{Validation}
\vspace*{-2mm}
\subsection{Validation of the models}\label{sec:validation}
\begin{figure}[!ht]
\centering\includegraphics[width=0.9\linewidth]{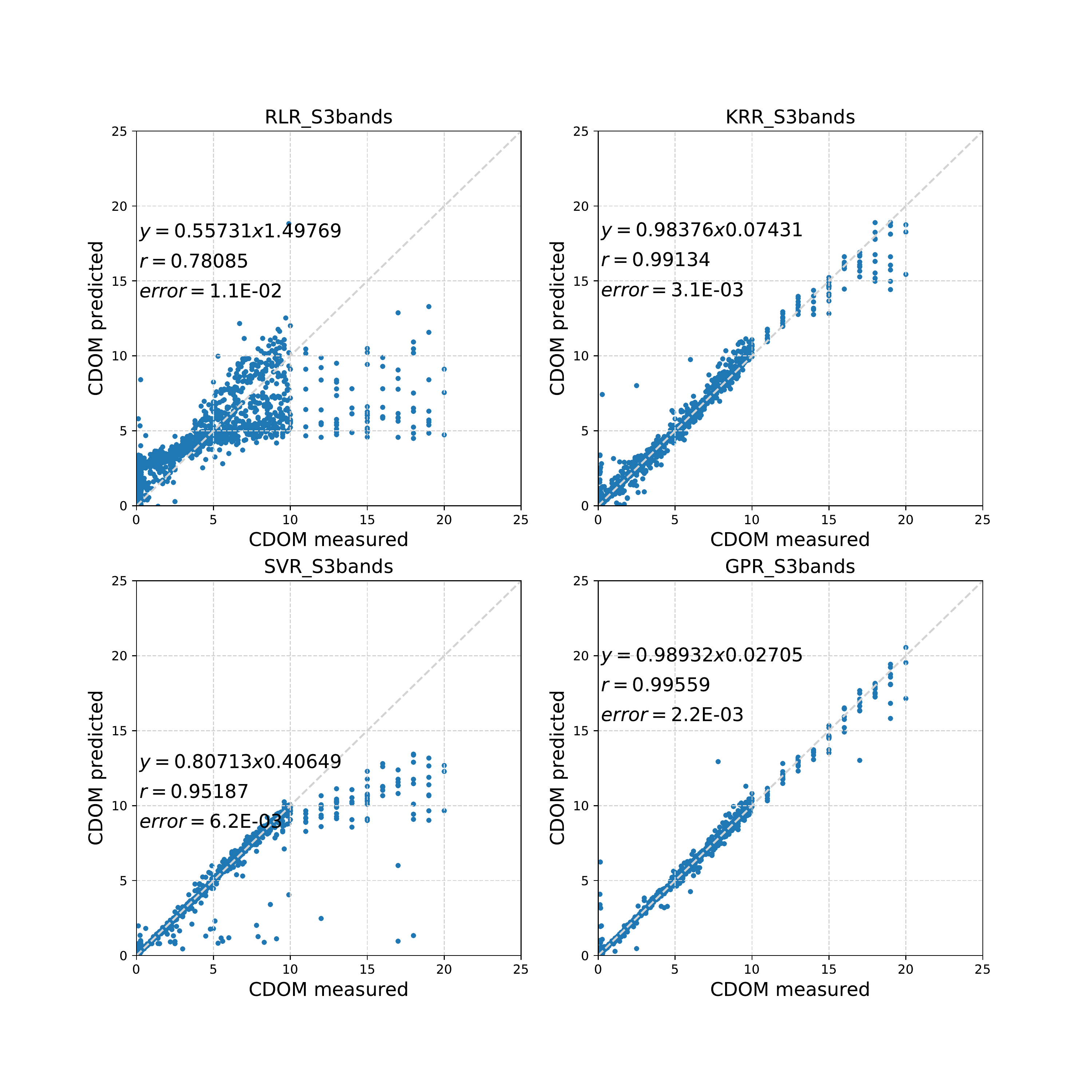}
\vspace*{-7mm}
\caption{\small Validation of the CDOM modeled against independent test data for the RLR, KRR, SVR and GPR models}\label{Fig3}
\end{figure}

\begin{figure}[!ht]
\centering
\includegraphics[width= 0.9 \linewidth]{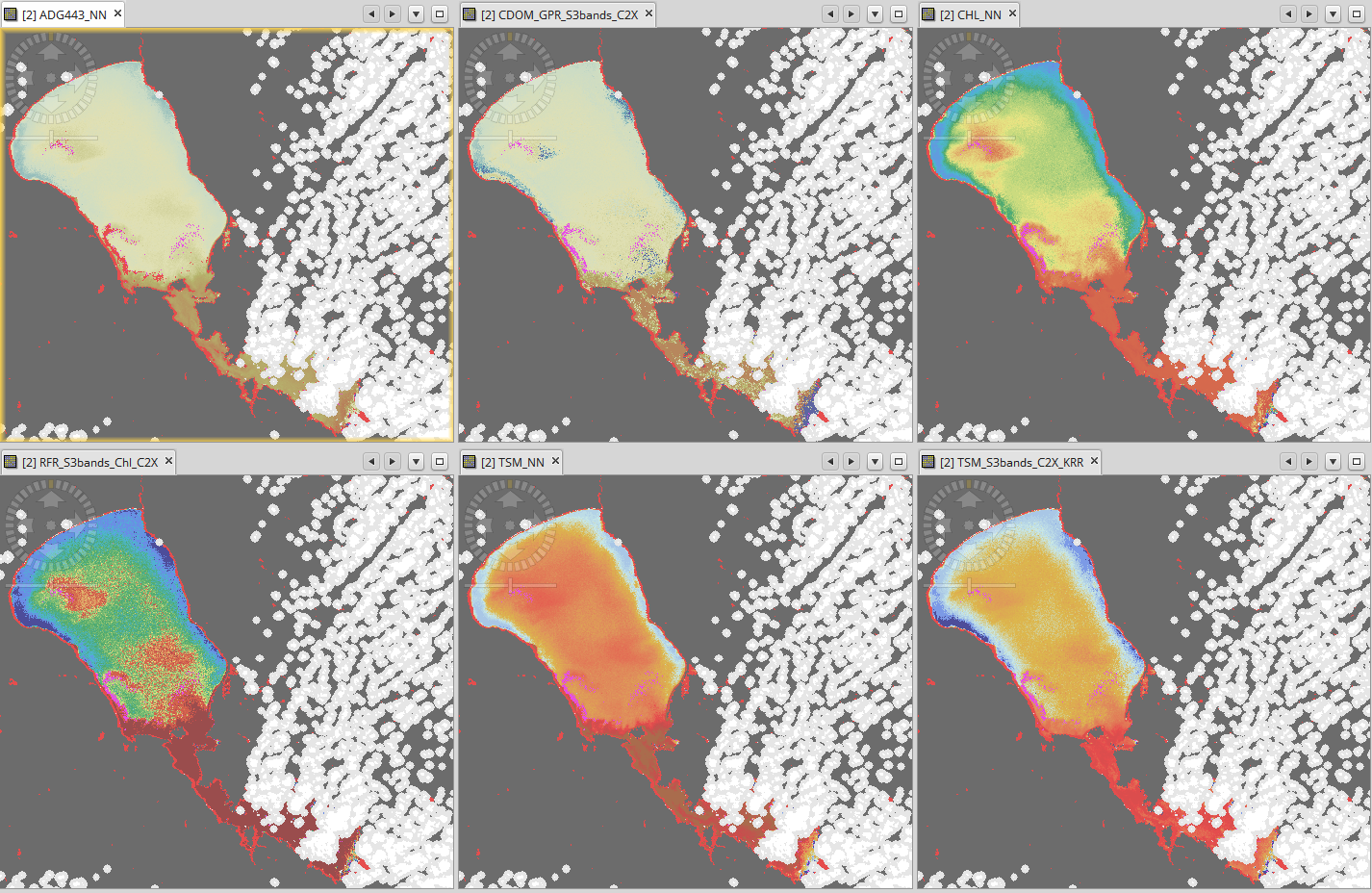}\\
\includegraphics[width= 0.3 \linewidth]{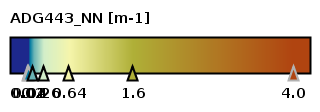}
\includegraphics[width= 0.3 \linewidth]{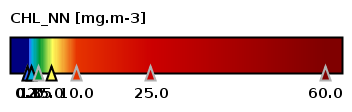}
\includegraphics[width= 0.3 \linewidth]{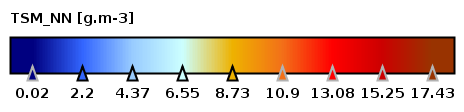}
\vspace*{-5mm}
\caption{\small Application to an OLCI scene (C2RCC $R_{rs}$), in gray the land pixels and in white the clouded areas. From left to right and top to bottom ADG443\_NN, CDOM\_GPR, CHL\_NN, CHL\_RFR, TSM\_NN, TSM\_KRR }\label{Fig4}
\end{figure}

\vspace*{-2mm}
As indicated before, the validation is done using an independent dataset. In general quite high coefficients of correlation are found ($>0.9$) for CDOM and TSM, except for the RLR method for CDOM (0.781). Chl-a shows worse results than the other two parameters, with a range between $0.7>r<0.86$ (see Table \ref{Tab:stats}), and higher standard errors. Again the RFR and KRR have a better adjustment. RFR and GPR are the two best models for CDOM, especially the GPR, with really small standard errors and an almost 1:1 adjustment between measured and estimated data. For the TSM all models work pretty well, with the RFR (with a slight positive bias) and the KRR and SVR with very similar and outstanding results, followed closely by the RLR (Figure \ref{Fig3}).\\
One example of the application of the models to an OLCI scene is shown in Figure \ref{Fig4}. The scene was downloaded from the CODA EUMETSAT service \footnote{\url{https://eoportal.eumetsat.int}} and the date of acquisition is 24 May 2016 (IPF-OL-2 06.11). We used the $R_{rs}$ bands from C2RCC algorithm \cite{Doerffer07} available in the Sentinel-3 Toolbox of the SeNtinel Application Platform (SNAP) as inputs for the models. We focused the analysis on the Peipsi lake in Estonia, but the scene maps a big part of the Baltic Sea and the Gulf of Finland. The \textit{ADG443\_NN} stays besides the CDOM GPR product, with very similar patterns and values. The \textit{CHL\_NN} shows bigger dissimilarities with the RFR Chl-a product in terms of values, but patterns are again very similar.\textit{TSM\_NN} is compared with the TSM KRR model and both products show similar patterns, with small differences mapped in the shores of the lake.In general, maximum differences occur in areas already identified as the ones with maximum uncertainties in the associated bands (not shown here).
\vspace*{-4mm}
\vspace*{-5mm}
%\vspace*{-5mm}
\subsection{Comparison with ONNS}\label{sec:comparison}
\vspace*{-2mm}
Results of the methods are also compared with the outputs of the ONNS (version 0.4), developed within the C2X project~\cite{Hieronymi2016}. ONNS uses a fuzzy logic classification scheme for selecting and blending optimized neural networks. Thirteen optical water types have been identified for selected OLCI wavebands based on $R_{rs}$. Three sets of NNs have been used to derive concentrations (Chl-a, CDOM, TSM), selected IOPs and other OC products. A direct comparison with the results shown in previous section is not possible since we merged the two water classes in one, and further filtered the data. However, attending at the statistics published in Table 3 of \cite{Hiero17}, we observe that the ONNS method gives results very similar to the ML approaches applied here: the Chl-a seems to be retrieved with more difficulty than the other two parameters (much higher RMSE and bias and lower \textit{r}), while CDOM and TSM show comparable good results with \textit{r} $>0.9$ and comparable low RMSE, slightly higher in the C2AX cases. 
\vspace*{-2mm}

%%%%%%%%%%%%%%%%%%%%%%%%%%%%%%%%%%%%%%%%%%%%%%%%%%%%%%%
\vspace*{-2mm}
\section{CONCLUSION}
\label{sec:con}
\vspace*{-2mm}
To overcome the disadvantages of the spatio-temporal limitations of the semi-empirical algorithms, five ML regression methods are tested using as training simulated $R_{rs}$ from the C2X project. The used simulated data represents medium to high absorbing waters with high CDOM concentrations, variable Chl-a content and relatively low TSM. The OLCI band configuration is used as basis. Results show that ML linear and non-linear regression methods are very efficient when using all available bands, adding information coming from different parts of the spectrum. For the three parameters (Chl-a, CDOM and TSM)retrieved, RLR is not the best in any case, but it is still good enough for TSM, which could be advantageous due to is easy implementation. For Chl-a and CDOM, RFR and GPR seem to be the best options, giving better or similar results than other more sophisticated approaches like C2RCC and ONNS. The possibility of making further tests with the other water types -complete C2X dataset-; and the use of other approaches like the deep GP are open lines of research.
% \section{REFERENCES}
% \label{sec:ref}

% List and number all bibliographical references at the end of the paper.  The references can be numbered in alphabetic order or in order of appearance in the document.  When referring to them in the text, type the corresponding reference number in square brackets as shown at the end of this sentence \cite{}.

%\scriptsize
%\vspace*{-2mm}
\newpage
\footnotesize
\bibliographystyle{IEEEbib}
\bibliography{WQML.bib}

\begin{thebibliography}{10}

\bibitem{Kallio183}
K.~Kallio,
\newblock ``Optical properties of {Finnish} lakes estimated with simple
  bio-optical models and water quality monitoring data,''
\newblock {\em Hydrology Research}, vol. 37, no. 2, pp. 183--204, 2006.

\bibitem{ATTILA2013138}
J.~Attila, S.~Koponen, K.~Kallio, M.~Lindfors, S.~Kaitala, and P.~Ylostalo,
\newblock ``{MERIS Case II} water processor comparison on coastal sites of the
  northern {Baltic Sea},''
\newblock {\em Remote Sensing of Environment}, vol. 128, no. Supplement C, pp.
  138 -- 149, 2013.

\bibitem{Hieronymi2016}
M.~Hieronymi, H.~Krasemann, D.~Mueller, C.~Brockmann, A.B. Ruescas, K.~Stelzer,
  B.~Nechad, K.~Ruddick, S.~Simis, G.~Tisltone, F.~Steinmetz, and P.~Regner,
\newblock ``Ocean colour remote sensing of extreme {Case-2} waters,''
\newblock in {\em Proceedings of the LPS}. Living Planet Symposium, 2016.

\bibitem{LIGI201757}
M.~Ligi, T.~Kutser, K.~Kallio, J.~Attila, S.~Koponen, B.~Paavel, T.~Soomets,
  and A.~Reinart,
\newblock ``Testing the performance of empirical remote sensing algorithms in
  the {Baltic Sea} waters with modelled and in situ reflectance data,''
\newblock {\em Oceanologia}, vol. 59, no. 1, pp. 57 -- 68, 2017.

\bibitem{Doerffer07}
R.~Doerffer and H.~Schiller,
\newblock ``The {MERIS Case 2} water algorithm,''
\newblock {\em International Journal of Remote Sensing}, vol. 28, no. 3-4, pp.
  517--535, 2007.

\bibitem{Hiero17}
M.~Hieronymi, D.~Mueller, and R.~Doerffer,
\newblock ``The {OLCI Neural Network Swarm (ONNS)}: A bio-geo-optical algorithm
  for open ocean and coastal waters,''
\newblock {\em Frontiers in Marine Science}, vol. 4, pp. 140, 2017.

\bibitem{Breiman85}
L.~Breiman and J.H. Friedman,
\newblock ``Estimating optimal transformations for multiple regression and
  correlation,''
\newblock {\em Journal of the American Statistical Association}, vol. 80, no.
  391, pp. 1580--598, 1985.

\bibitem{Shawetaylor04}
J.~Shawe-Taylor and N.~Cristianini,
\newblock {\em Kernel Methods for Pattern Analysis},
\newblock {Cambridge University Press}, 2004.

\bibitem{Rasmussen06}
C.~E. Rasmussen and C.~K.~I. Williams,
\newblock {\em Gaussian Processes for Machine Learning},
\newblock The MIT Press, New York, 2006.

\bibitem{Ruescas18}
A.B. Ruescas, M.~Hieronymi, G.~Mateo-Garcia, , S.~Koponen, K~Kallio, and
  G.~Camps-Valls,
\newblock ``Machine learning regression approaches for colored dissolved
  organic matter ({CDOM}) retrieval with {S2-MSI} and {S3-OLCI} simulated
  data,''
\newblock {\em Remote Sensing}, p.~24, 2018, submitted.

\end{thebibliography}

\end{document}